\documentclass[12pt, onecolumn,article]{IEEEtran}

\usepackage[utf8]{inputenc}
\usepackage{cite}
\usepackage{setspace}
\usepackage{epsf}\epsfverbosetrue
\usepackage{graphics,epsfig}
\usepackage{graphicx,epsfig}
\usepackage{epsfig}
\usepackage{multirow}
\usepackage{alltt}
\usepackage{amsfonts}
\usepackage{subfigure}
\usepackage{tcolorbox}
\usepackage{float}
\usepackage{url}
\usepackage{graphicx}
\usepackage{verbatim} 
\usepackage{footnote} 
\usepackage{latexsym} 
\usepackage{sidecap} 
\usepackage{wrapfig}
\usepackage{eso-pic}
\usepackage{fix-cm}
\usepackage{algorithm}
\usepackage{algorithmic}
\usepackage{color}
\usepackage{wrapfig}
\usepackage{multicol}

\usepackage{amsmath}
\usepackage[
  locale = DE 
]{siunitx}

\usepackage{textcomp}
\usepackage{multirow}
\usepackage{mathtools}
\usepackage{soul} 
\usepackage{amsmath} 
\usepackage{verbatim}
\usepackage{csvsimple} 
\usepackage{array}
\usepackage{subfig}
\usepackage[normalem]{ulem}
\usepackage{booktabs}
\newcolumntype{P}[1]{>{\RaggedRight\arraybackslash}p{#1}}
\newcommand{\tabitem}{\textbullet~~}

\usepackage{pifont}
\newcommand{\tick}{\ding{52}}%
\newcommand{\cross}{\ding{55}}%
\begin{document}

\title{Performance Evaluation of Differential Privacy Mechanisms in Blockchain based Smart Metering}

\author{Muneeb Ul Hassan, Mubashir Husain Rehmani, and Jinjun Chen
\thanks{M. Ul Hassan and J. Chen are with the Swinburne University of Technology, Hawthorn VIC 3122, Australia  (e-mail:  muneebmh1@gmail.com; jinjun.chen@gmail.com).}
\thanks{M.H. Rehmani is with the Department of Computer Science, Cork Institute of Technology, Rossa Avenue, Bishopstown, Cork, Ireland (e-mail: mshrehmani@gmail.com).}
}

\maketitle

The concept of differential privacy emerged as a strong notion to protect database privacy in an untrusted environment. Later on, researchers proposed several variants of differential privacy in order to preserve privacy in certain other scenarios, such as real-time cyber physical systems. Since then, differential privacy has rigorously been applied to certain other domains which has the need of privacy preservation. One such domain is decentralized blockchain based smart metering, in which smart meters acting as blockchain nodes sent their real-time data to grid utility databases for real-time reporting. This data is further used to carry out statistical tasks, such as load forecasting, demand response calculation, etc. However, in case if any intruder gets access to this data it can leak privacy of smart meter users. In this context, differential privacy can be used to protect privacy of this data. In this chapter, we carry out comparison of four variants of differential privacy (Laplace, Gaussian, Uniform, and Geometric) in blockchain based smart metering scenario. We test these variants on smart metering data and carry out their performance evaluation by varying different parameters. Experimental outcomes shows at low privacy budget ($\varepsilon$) and at low reading sensitivity value ($\delta$), these privacy preserving mechanisms provide high privacy by adding large amount of noise. However, among these four privacy preserving parameters Geometric parameters is more suitable for protecting high peak values and Laplace mechanism is more suitable for protecting low peak values at ($\varepsilon$ = 0.01).

\section{Introduction}
Differential privacy have been proposed by C. Dwork in 2006, and since then it has been applied to various domains to protect their privacy~\cite{intref01}. Differential privacy provides an insightful privacy definition which can be used to perturb raw data records by adding an adequate amount of noise drawn from respective distribution~\cite{intref02}. Informally, the notion of differential privacy guarantees that addition, modification, deletion, or variation of a single record within a dataset will not have any significant effect on the output query results~\cite{intref03}. Initially, differential privacy notion was proposed to protect privacy of statistical databases, however later experiments showed that it can effectively be applied to other real-life scenarios as well, such as real-time reporting and machine learning. Afterwards, researches are being carried out to apply differential privacy in major domains that require privacy protection including smart grid, cloud computing, industries, and other similar cyber physical systems~\cite{hassan01, jinref01}.\\
Conventional smart grid networks did not used any specific security of privacy strategies and just used to rely on security and privacy provided by communication protocols, however, with the passage of time, it was found out that data of smart grid can be used to carry out major privacy and security breaches~\cite{intref05, jinref03}. Since then, plenty of researches are being carried out to enhance smart grid technology by improving its security and privacy. For example, the use of homomorphic cryptosystem and trusted remote entity have been proposed by researchers to overcome these eavesdropping issues~\cite{intref04, jinref02}. One such way that effectively enhances the security and trust of smart grid network is the integration of blockchain in smart grid domain~\cite{intref06, jinref05}. Blockchain network ensures that all the communication and storage being carried out via blockchain network is secure and adversaries will not be able eavesdrop into the privacy of blockchain users. This is done by using advanced secure technologies, such as cryptographic hashing, tamper-proof record storing, and strong distributed consensus~\cite{intref07}. \\
Despite of this secure nature, it has been highlighted that blockchain based smart grid network is still vulnerable to certain privacy threats because of its decentralized nature~\cite{hassan02}. For example, an adversary can compromise a specific smart meter after analysing the available data on decentralized blockchain ledger. Similarly, the stored data on grid utility database can be analysed to infer into private information of consumer usage patterns. In order to mitigate such issue, certain researches have been carried out that involved integration of certain privacy preservation approaches in blockchain scenarios, such as zero-knowledge proofs and anonymization~\cite{hassan02, jinref04}. These works are viable to certain extent, but they cannot directly be applied to real-time smart metering networks because majority of them either works over stored data or only works over private data provenance. One such mechanism that can effectively protect privacy of smart grid users in a decentralized blockchain scenario is differential privacy~\cite{hassan04}. Differential privacy can protect this information because of its dynamic nature, especially point-wise perturbation mechanism of differential privacy can protect real-time data without running extensive computationally complex algorithms~\cite{intref08}.\\
In this chapter, we work over integration of differential privacy protection mechanism in decentralized blockchain based smart metering scenario. To examine it further, we evaluate four variants of differential privacy (Laplace, Gaussian, Uniform, and Geometric) on real-time smart metering data. In order to check their efficiency and effectiveness we use mean absolute error (MAE) as evaluation parameter. Experimental results demonstrate that each mechanism has their own pros and cons depending upon the privacy budget, sensitivity value, and the data applied over it. For instance, all mechanisms provide high level of privacy when the privacy parameters ($\varepsilon \& \delta$) have low values (e.g., $\varepsilon$ = 0.01 \& $\delta$ = 0.01). However, when these values increases, the privacy of the meter reading value reduces gradually and at $\varepsilon$ = 1 \& $\delta$ = 1, the privacy reaching to a minimum level, although data utility is maximum at this stage. Similarly, among these four privacy notions, Geometric and Laplace performs better at lower privacy budgets by adding sharp amount of noise. Specifically,  when there are high peaks values in metering data (e.g., high usage), then the Geometric mechanism preserves privacy is the most proficient manner, and when the smart metering data has low peaks (e.g., less occupancy/usage), Laplace mechanism outperforms other mechanisms. 
\subsection{Key Contributions}
The key contributions of our book chapter are as follows:
\begin{itemize}
\item We integrate differential privacy in blockchain based smart metering scenario.
\item We carry out in-depth performance evaluation of differential privacy mechanisms in decentralized blockchain scenarios at different privacy budget ($\varepsilon$) values.
\item From experimental results, we analyse the effectiveness of variants of differential privacy along with the reported mean absolute error rate. We conclude that Geometric noise addition mechanism outperforms other mechanism in high peak values, however, for low peak values, Laplace mechanism outperforms other mechanisms.
\end{itemize}


\begin{table*}[t!]
\begin{center}
 \centering
 \footnotesize
 \captionsetup{labelsep=space}
 \captionsetup{justification=centering}
 \caption{\textsc{\\Comparative View of Works Carried out in Smart Metering from Perspective of Differential Privacy (DP) and Blockchain.}}
  \label{tab:sgtab01}
  \begin{tabular}{|P{1.5cm}|P{0.55cm}|P{2cm}|P{2cm}|P{0.95cm}|P{0.95cm}|P{1.3cm}|}
  	\hline
\rule{0pt}{2ex}
\bfseries \centering Name of Strategy & \bfseries \centering Ref No. & \bfseries \centering Major Contribution & \bfseries \centering Parameters Enhanced & \bfseries \centering Consid-\newline ered DP & \centering \bfseries Consid-\newline ered Block-\newline chain & \bfseries Simulator Used  \\
\hline

\multirow{3}{*}{\parbox{2cm}{\centering \textbf{}}}

\rule{0pt}{2ex}
DREAM & ~\cite{intref09} &  Introduced the concept of DP in Smart Metering & \tabitem Appliance privacy protection & \centering \tick & \centering \cross &  Electricity Trace Simulator \\
\cline{1-7}

\rule{0pt}{2ex}
DP for Real Smart Metering Data & ~\cite{intref10} &  Efficient DP mechanism to balance utility-privacy & \tabitem Aggregated data protection via smoothing & \centering \tick & \centering \cross & $N/A$  \\
\cline{1-7}

\rule{0pt}{2ex}
DP for RER based Smart Metering & ~\cite{intref11} &  Protected usage and generation privacy of RER based smart homes & \tabitem Peak-load protection \newline \tabitem RER generation protection & \centering \tick & \centering \cross &  Python  \\
\cline{1-7}

\rule{0pt}{2ex}
GridMonitoring & ~\cite{intref12} &  Monitoring smart grid values via blockchain & \tabitem Enhanced provenance \& transparency \newline \tabitem Enhanced trust & \centering \cross & \centering \tick &  $N/A$  \\
\cline{1-7}

\rule{0pt}{2ex}
Light-weight blockchain based AMI & ~\cite{intref13} &  AMI network is protected and secured via blockchain & \tabitem Enhanced received signal strength & \centering \cross & \centering \tick & MATLAB \\
\cline{1-7}

\rule{0pt}{2ex}
Blockchain based Secure Smart Grid & ~\cite{intref14} &  Key-less secure signature scheme for decentralized smart grid & \tabitem Developed automated access control manager & \centering \cross & \centering tick & GoEth  \\
\cline{1-7}

\rule{0pt}{2ex}
DP Variants in decentralized Smart Metering & This Work &  Performance evaluation of DP variants in blockchain based smart metering & \tabitem Comparison of DP variants at different privacy parameters \newline \tabitem MAE comparison & \centering \tick & \centering \tick & Python \\
\cline{1-7}

\hline

 \end{tabular}
  \end{center}
\end{table*}


\subsection{Related Work}
Since the advent of modern smart grid, researches are being carried out to make it autonomous, secure, and user friendly. In order to do so, plenty of works targeted privacy preservation or integration of blockchain in smart grid scenario. For example, the first work that highlighted the use of differential privacy in smart metering scenario was carried out by Acs~\textit{et al.} in~\cite{intref09}. The presented work in the article used gamma distribution based differential privacy protection to preserve smart metering data and also used encryption based cryptography to enhance security during transmission. Authors also worked over the phenomenon of multi-slot privacy in which they effectively use the dynamic nature of differential privacy to protect this real-time data. Another similar work was carried out by Eibl~\textit{et al.} in ~\cite{intref10}. The authors for the first time discussed the terminology of point-wise differential privacy protection for real-time smart data. Authors evaluated the use of Laplace noise and worked over signal smoothing to reduce the risk for privacy leakage in case of any adversarial attack. Similarly, a work that targets the integration of differential privacy in renewable energy resources have been discussed by authors in~\cite{intref11}. The work discussed the protection of energy being generated from their resources by adding dynamic Laplace noise. The work also introduced the concept of peak protection in renewable energy resources reporting to ensure that privacy of smart meter users remain unviolated despite of any adversarial attack on grid database.\\
Similarly, from perspective of blockchain based smart grid, a work that discusses the integration of blockchain in grid monitoring scenario was carried out by Gao~\textit{et al.} in~\cite{intref12}. Authors in this work used blockchain to ensure transparency and to enhance trust in the network by providing information publicly available to users via decentralized distributed ledger. Furthermore, authors provide a platform to users via which they can monitor their usage without depending upon any third-party. One more work that discusses integration of blockchain in advance metering infrastructure (AMI) to enhance its security and transparency is carried out by Kamal~\textit{et al.} in~\cite{intref13}. The major motto of the work is to protect smart meters from various cyberattacks specifically targeting data tampering, and man-in-the-middle attacks. Authors did this by proposing a light-weight blockchain based solution for decentralized AMI network. Similarly, authors in~\cite{intref14} proposed a keyless blockchain based signature scheme for smart grid network via which they enhance security of traditional smart grid. Authors also discussed the aspect of trusted third party breaches and failures and suggested that use of blockchain based secure platform is a viable solution to overcome these issues. Furthermore, authors claimed that the proposed strategy turns blockchain network into an automated manager to carry out access-control operation on smart grid. Another significant contribution carried out by authors is the enhancement of storage cost over the decentralized blockchain network, which is one of the major issue blockchain is facing nowadays. \\
After analysing all these works, we can say that to the best of our knowledge, no work that discuss the integration of differential privacy in blockchain based smart metering network have been carried out in the literature. In this chapter, we not only discuss the integration of differential privacy in decentralized smart metering, but also evaluate four major mechanisms of differential privacy in this network to check their effectiveness. 

\subsubsection{Organization of This Chapter}
The remainder of this chapter is organized as follows:
Section 1.2 discusses preliminaries of our work including differential privacy, blockchain, and real-time smart metering, Section 1.3 provided detailed discussion about system model, differentially private reporting algorithms, along with design goals and adversary model. Furthermore, Section 1.4 provides performance evaluation of variants of differential privacy in blockchain based smart metering scenario. Finally, the work is concluded by providing conclusion and future directions in Section 1.5.

\section{Preliminaries of Our Work}
In this section, we discuss the preliminaries involved in our evaluation, ranging from differential privacy mechanisms to blockchain and smart grid.
\subsection{Differential Privacy Mechanisms} \label{DPVar}
Differential privacy can be termed as a notion to protect privacy in an adversarial environment~\cite{newref01}. Formally, differential privacy can be defined as a randomize response that ensures that query evaluation of two neighbouring datasets varying by just one element will produce similar output results that will introduce randomness in the results~\cite{newref02}. The equation is as follows:
\begin{equation}
P_r[F_{(db_1)}) \in R] \leq \exp(\varepsilon) \times P_r[F_{(db_2)}) \in R] + \delta
\end{equation}

In above equation, $\varepsilon$ is the privacy budget that controls the amount of noise being added. $\delta$ is the sensitivity value that is usually determined on the basis of dataset. $R$ is the query output range for query function $F$. 
Similarly, the formula of sensitivity calculation for  two adjacent datasets (${db_1}$) \& (${db_2}$) can be defined as follows~\cite{hassan01,newref03}:

\begin{equation}
\label{seneqn}
\Delta F_s = \max\limits_{db_1 , db_2} ||F(db_1) – F(db_2)||
\end{equation}

Apart from the formal definitions, researchers worked over proposing of various variants of differential privacy to support different privacy needs. These variants can be classified into distribution protection mechanisms and data perturbation mechanisms. In this section, we discuss four major data perturbation mechanisms named as Laplace, Gaussian, Uniform, and Geometric mechanisms of differential privacy. A detailed discussion about notions and variants of differential privacy can be found in~\cite{preref01}. 
\subsubsection{Laplace Mechanism}
Laplace mechanism is considered to be the pioneering mechanism which was used by C. Dwork at the time of proposal of this notion of differential privacy perturbation. Afterwards, this notion has been used widely to protect privacy at different application scenarios~\cite{hassan05}. Taking insights from sensitivity equation (Eqn.~\ref{seneqn}), we can demonstrate Laplace noise by considering $S_c = \frac{\Delta F_s}{\varepsilon}$  as follows~\cite{preref02}:

\begin{equation}
Lap(S_c) = \exp{(-\frac{|x-\mu|}{S_c})}
\end{equation}

Since, the standard deviation of the given function is calculated using an exponential distribution which is symmetric with respect to parameter $\sqrt{2}b$. Then, the probability density function (pdf) of traditional Laplace noise at mean value ($\mu$) is defined as follows:

\begin{equation}
\label{lapeqn}
Pdf(x) = \frac{\exp{(\frac{|x-\mu|}{b})}}{2b}
\end{equation}
The noise is then computed via probability density function, and then added to the query/point-wise reading result in order to protect privacy.

\subsubsection{Gaussian Mechanism}
Gaussian mechanism also known as bell curve distribution has also been widely used as a traditional notion of differential privacy. The bell shape of Gaussian probability distribution provided fine-grained noise that can be added to query evaluation in order to integrate randomness in the output results~\cite{hassan01}. The standard formula for Gaussian mechanism is as follows:
\begin{equation}
\label{gaus}
pdf(x) = \frac{1}{\sqrt{ 2 \pi b^2 }} e^{ - \frac{ (x - \mu)^2 } {2 b^2} }
\end{equation}
In the above equation, $b$ or the standard deviation is used as a scale to choose the appropriate amount of noise, which is controlled by the privacy budget value $\varepsilon$.

\subsubsection{Uniform Mechanism}
The concept of using uniform distribution as a notion of differential privacy have been proposed by plenty of researchers because of its strong privacy along with low computational complexity. For example, Kalantari~\text{et al.} analysed privacy-utility trade-offs via uniform differential privacy and integrated this concept with hamming distortion in~\cite{preref03}. Similarly, Geng and Viswanath discussed it under the umbrella of optimal noise adding mechanisms in their article~\cite{preref04}. \\
Formally, uniform distribution noise is a discrete noise addition mechanism which works over sensitity value (0, $\delta$), instead of just being dependent upon $\varepsilon$. Noise in this mechanism is computed using uniform probability distribution as follows:
\begin{equation}
\label{uniform}
pdf(x) = \begin{cases}
      \frac{\delta}{\Delta}, & \text{if}\ \forall - \frac{\Delta}{2\delta} \leq k \leq \frac{\Delta}{2\delta}  -1  \\
      0, & \text{otherwise}
    \end{cases}
\end{equation}

\subsubsection{Geometric Mechanism}
Geometric differential privacy mechanism was also proposed by researchers under the umbrella of oblivious notions of differential privacy. This notion is pretty diverse and plenty of sub-mechanism/variants have been proposed that constitute of various ways to draw differentially private noise from Geometric distribution. For example, authors in~\cite{preref05} discussed it as a discrete version of traditional Laplace mechanism and states that it generates optimal utility value for all type of Bayesian information counting queries. Similarly, Ghosh~\textit{et al.} proposed the notion of Geometric, and truncated Geometric differential privacy noise addition to perturb query results~\cite{preref06}. Formally, Geometric mechanism  of differential privacy can be defined as follows:

\begin{equation}
pdf(X=x) = \frac{1-\alpha}{1+\alpha}\alpha^{|x|}, \forall x \in \mathbb{Z}
\end{equation}

\subsection{Real-Time Smart Metering and Privacy Issues}
Smart meter is a device that serves as a bridge between grid utility and smart homes as it links grid utility with smart home via strong communication medium. Together, all smart meters within a range constitute a network named as advanced metering infrastructure (AMI)~\cite{preref07}. In order to carry out various statistical processes such as net metering, demand response calculation, load forecasting, and load scheduling, these smart meters send their actual real-time fine grained meter reading to grid utility. These values are then stored int their databases which can be accessed for query evaluation after getting approval from authorities. It definitely is a good model to carry out statistical analysis, however, one of the major drawback of this model is that it leaks privacy of smart meter users~\cite{preref08}. For example, the reported fine-grained data from smart meters can further be fed to various non-intrusive load monitoring (NILM) algorithms that can even provide information about usage of any specific appliance at a specific time by visualizing load curves~\cite{preref09}. \\
This information can also lead to carry out various criminal/ burglar activities at the time of unoccupancy of house. Therefore, it is important to protect this information via some privacy preserving mechanism. 

\subsection{Blockchain Network}
Blockchain came into limelight after successful functioning of decentralized cryptocurrency called Bitcoin by Satoshi Nakamoto~\cite{preref10}. Formally, blockchain is categorized into two major types from perspective of permission named as permissioned blockchain and permission-less blockchain. Furthermore, from perspective of availability and controlling authorities, it is categorized into three major types called as public, private, and consortium. A detailed discussion about these type and their pros and cons are available in~\cite{preref11}. In this chapter, we use public permissionless blockchain in which every smart meter node can join after filling a detailed form to ensure its legitimacy. Since, it is a public blockchain, so every blockchain node can participate in the consensus part and can earn extra reward by mining the block. Since smart meters have less computational power, therefore they can only take part in consensus if the provided consensus mechanism is not dependent upon computation power. In order to fulfil this requirement, we use proof-of-stake (PoS) instead of traditional computationally expensive proof-of-work (PoW) consensus mechanism. \\
In our PoS mechanism, every smart meter acting as a blockchain node can take part in the mining process after depositing a specific number of tokens in the network. This eradicates the need for having high mining power. Similarly, in this way smart meters can also incentivize themselves a bit more if they are interested to contribute to the network. The incentive can be in the form of mining reward that a mining node will get if it gets selected as a winning miner.

\section{Functioning and System Model}
In this section, we present system model, design goals, adversary model, and algorithmic foundation of integration of differential privacy in blockchain based smart metering scenario.
\subsection{System Model}
\begin{figure*}[t]        
\includegraphics[scale = 0.5]{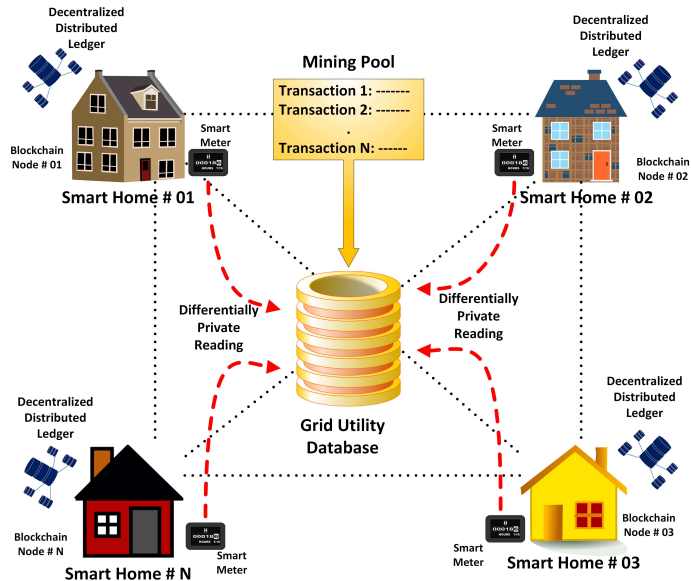}
\centering
  \caption{An Illustrative Demonstrating of Our Proposed System Model for Differentially Private Blockchain-based Smart Metering}     
  \label{fig:sysmod}   
\end{figure*}

Our application scenario  consists of two important entities, i.e., smart homes (deployed with smart meters) and grid utility. Our scenario works over the model of a public blockchain, which means that all the participating nodes can take part in the mining process. The most important entity in these nodes are smart homes that contains smart meters which are reporting their real-time energy to grid utilities. Each smart meter is a blockchain node which can take part in the mining process after adding some tokens as a stake in the network. A detailed system model is given in Fig.~\ref{fig:sysmod}.\\
From Fig.~\ref{fig:sysmod}, it can be seen that each smart home is connected to every other smart home via decentralized blockchain network, similarly, these smart homes are also connected to the grid utility via same network. However, an additional feature which distinguish this specific connection from other is that these smart homes regularly report their real-time values to grid database utility after the specified time interval (e.g., 10 minutes). Grid utility collect these values from all smart homes and put these values along with some other transaction values in a public mining pool. The values at mining pool can be viewed publicly by all mining nodes in order to facilitate them in mining the block. \\
On the other hand, all miner nodes (smart homes) participate in the mining process by putting some stakes in the network. In PoS mining, a miner is choses in accordance with the sake it has invested in the network. For example, a node which have invested 50\% stakes as compared to all other nodes have approximately 50\% chances of being selected as the miner. Contrary to this, if the proportion of stakes of some node is even less than 1\% then its chances of winning the mining election is fairly low. Coming back to the mining process; these miner nodes submit their stakes and wait for the mining process to select miners in the basis of stake probability. The mechanism selects a miner node according to the mentioned process, and the selected miner then move further towards the next process.\\
Once the miner gets finalized, it selects all the available transaction from the mining pool, compute their hash, and form a block like structure. After formation of block, this miner disseminates the block to all other blockchain nodes via broadcasting mechanism in order to get verification votes. All receiving nodes compute hash at their end and verify the content of block. Once the verification stage gets completed, the block is then added to the blockchain network and is disseminated as a mined block. In this way, each blockchain node will have a copy of updated decentralized ledger that can be backtracked at any time to ensure transparency and enhance trust in the network.\\
This complete process involves dealing with real-time reported energy values, therefore, protecting the privacy of these reading before disseminating in mining pool, or for verification is pretty important in order to overcome any catastrophe. Therefore, in our proposed model we protect these readings via adding differentially private noise before transmitting them to grid utility for evaluation. A graphical representation of noise addition via smart homes is provided in Fig.~\ref{fig:sysmod}.

\subsection{Design Goals}
Previous work in privacy protection in real-time smart metering do not incorporate the advantageous nature of blockchain and differential privacy at the same time. Some previous works evaluated the use of differential privacy in smart metering, while some works discussed the integration of blockchain with smart metering. However, no combined literature that integrates all three technologies of smart metering, blockchain, and differential privacy have been discussed by researchers. The design goals of our proposed scenario are as follows:
\begin{itemize}
\item Integrating public blockchain network in a smart metering networks using the concept of decentralized distributed ledger.
\item Enhancing security of smart metering network by providing a model to carry out distributed consensus in the network.
\item Maintaining trust and transparency in the network by updating the decentralized ledger after addition of every new block.
\item Ensuring privacy of real-time smart metering data in run-time by adding dynamic differentially private noise in the data.
\item Enhancing mean absolute error of the network by carrying out comparison between differential privacy mechanisms.
\end{itemize}

\subsection{Adversary Model}
Adversaries in our scenario can be any type of intruders that could be interested to get information regarding real-time usage of smart meter electricity. This could be done in order to get more precise information regarding availability of occupants of smart homes so that they can carry out malicious/burglar activities. Moreover, some adversaries do also require fine-grained data to find out type of appliances being used in smart homes, or to know about appliances who are expected to get damaged in near future, etc. This is done by certain marketing companies and they use this fine-grained data to carry out targeting advertising~\cite{intref11}. Similarly, man-in-the-middle attack can also be carried out via which the information being transmitted from smart meter to grid utility can be misused after analyzation~\cite{sysref01}. Furthermore, data linking attack can also be carried out over the stored data in smart grid database, in which the data can be exploited and can be linked with certain other databased to get to know more information regarding any specific person~\cite{sysref02}. \\
In order to overcome these attacks and adversarial behaviours, we first integrate blockchain that ensure security and trust in the network via its cryptographic mechanisms. Furthermore, we work over integration of variants of differential privacy that enhances privacy of the complete network.


\begin{algorithm}[]\scriptsize
\caption{Algorithm for Selection of Differential Privacy Noise Variants}
\label{algo1}




\begin{algorithmic}[1]


\STATE \textbf{main()}

\STATE \hskip 1.5em $\varepsilon \gets$ Epsilon Value from User
\STATE \hskip 1.5em $\delta \gets$ Reading Sensitivity Value from User
\STATE \hskip 1.5em $M_R \gets$ Actual Meter Reading from Smart Meter
\STATE \hskip 1.5em $Func \gets$ Select Noise Function from User

\STATE \hskip 1.5em \textbf{Switch} (\textit{$Func$}) \textbf{do}
\STATE \hskip 3em \textbf{Case} \textit{Lap}\textbf{:}
\STATE \hskip 4.5em \textbf{Call} \textit{LaplaceNoiseMechanism($\varepsilon, M_R$)}
\STATE \hskip 4.5em \textbf{Break;}

\STATE \hskip 3em \textbf{Case} \textit{Gaus}\textbf{:}
\STATE \hskip 4.5em \textbf{Call} \textit{GaussianNoiseMechanism($\varepsilon, M_R$)}
\STATE \hskip 4.5em \textbf{Break;}

\STATE \hskip 3em \textbf{Case} \textit{Unif}\textbf{:}
\STATE \hskip 4.5em \textbf{Call} \textit{UniformNoiseMechanism($\varepsilon, \delta, M_R$)}
\STATE \hskip 4.5em \textbf{Break;}

\STATE \hskip 3em \textbf{Case} \textit{Geo}\textbf{:}
\STATE \hskip 4.5em \textbf{Call} \textit{GeoNoiseMechanism($\varepsilon,M_R$)}
\STATE \hskip 4.5em \textbf{Break;}

\STATE \hskip 1.5em \textbf{End Switch}

\STATE \textbf{end main()}

\end{algorithmic}
\end{algorithm}



\begin{algorithm}[H]\scriptsize
\caption{Differential Privacy Variants in Decentralized Smart Metering}
\label{algoDP}

 \textbf{Input:}  \textbf{N}, \textbf{F}, \textbf{$M_R$}, \textbf{$\varepsilon$}, \textbf{$\Delta$}, \textbf{$\delta$}

\textbf{Output:} \textbf{$P_V$}, \textbf{$A_E$}

\vskip 2mm
\hspace*{4mm} (1) \underline{Laplace Differential Privacy Mechanism}
\vskip 1mm

\begin{algorithmic}[1]


\IF{LaplaceNoiseMechanism()}
\FOR {\texttt{(each \textbf{j} in \textbf{N})}}

\STATE $\varepsilon_L \gets$ Laplace Privacy Budget
\STATE $\Delta \gets$ Database Sensitivity
\STATE \textbf{Generate} Laplace Noise via
\vskip 0.7mm
~\hskip 4.5em $\underset{j}{noise}$ = $Lap(F, \varepsilon_L, M_R)$
\vspace{-0.1em}
\STATE \textbf{Calculate} $\underset{j} {P_V} = \underset{j}{MR} + \underset{j}{noise}$
\vspace{-0.3em}
\STATE \textbf{Call} $AE_{function}(\underset{j} {P_V}, \underset{j}{M_R})$
\vspace{-0.7em}
\RETURN $\underset{j} {P_V}, \underset{j}{A_E}$
\vspace{-0.8em}
\ENDFOR
\ENDIF



\vskip 2mm
\hspace*{4mm} (2) \underline{Gaussian Differential Privacy Mechanism}
\vskip 1mm

\IF{GaussianNoiseMechanism()}
\FOR {\texttt{(each \textbf{j} in \textbf{N})}}

\STATE $\varepsilon_G \gets$ Gaussian Privacy Budget
\STATE $\Delta \gets$ Database Sensitivity
\STATE \textbf{Generate} Gaussian Noise via
\vskip 0.7mm
~\hskip 4.5em $\underset{j}{noise}$ = $Gaussian(F, \varepsilon_G, M_R)$
\vspace{-0.1em}
\STATE \textbf{Calculate} $\underset{j} {P_V} = \underset{j}{MR} + \underset{j}{noise}$
\vspace{-0.3em}
\STATE \textbf{Call} $AE_{function}(\underset{j} {P_V}, \underset{j}{M_R})$
\vspace{-0.7em}
\RETURN $\underset{j} {P_V}, \underset{j}{A_E}$
\vspace{-0.8em}
\ENDFOR
\ENDIF



\vskip 2mm
\hspace*{4mm} (3) \underline{Uniform Differential Privacy Mechanism}
\vskip 1mm

\IF{UniformNoiseMechanism()}
\FOR {\texttt{(each \textbf{j} in \textbf{N})}}

\STATE $\varepsilon_U \gets$ 0
\STATE $\Delta \gets$ Database Sensitivity
\STATE $\delta_U \gets$ Meter Reading Sensitivity Value
\STATE \textbf{Generate} Uniform Noise via
\vskip 0.7mm
~\hskip 4.5em $\underset{j}{noise}$ = $Uniform(F, \varepsilon_U, \delta_U, M_R)$
\vspace{-0.1em}
\STATE \textbf{Calculate} $\underset{j} {P_V} = \underset{j}{MR} + \underset{j}{noise}$
\vspace{-0.3em}
\STATE \textbf{Call} $AE_{function}(\underset{j} {P_V}, \underset{j}{M_R})$
\vspace{-0.7em}
\RETURN $\underset{j} {P_V}, \underset{j}{A_E}$
\vspace{-0.8em}
\ENDFOR
\ENDIF



\vskip 2mm
\hspace*{4mm} (4) \underline{Geometric Differential Privacy Mechanism}
\vskip 1mm

\IF{GeometricNoiseMechanism()}
\FOR {\texttt{(each \textbf{j} in \textbf{N})}}

\STATE $\varepsilon_{geo} \gets$ Geometric Privacy Budget
\STATE $\Delta \gets$ Database Sensitivity
\STATE $P_R \gets$ Probability of Success via $\varepsilon_{geo} \& \delta$

\STATE \textbf{Generate} Geometric Noise via
\vskip 0.7mm
~\hskip 4.5em $\underset{j}{noise}$ = $Geometric(F, P_R)$
\vspace{-0.1em}
\STATE \textbf{Calculate} $\underset{j} {P_V} = \underset{j}{MR} + \underset{j}{noise}$
\vspace{-0.3em}
\STATE \textbf{Call} $AE_{function}(\underset{j} {P_V}, \underset{j}{M_R})$
\vspace{-0.7em}
\RETURN $\underset{j} {P_V}, \underset{j}{A_E}$
\vspace{-0.8em}
\ENDFOR
\ENDIF


\vskip 1.2mm
(5)~\underline{Absolute Error Calculation Function}
\vskip 1mm

\STATE \textbf{Function} $AE_{function} ()$ \textbf{do}
\STATE ~\hskip 1.5em $M_R \gets$ Meter Reading
\STATE ~\hskip 1.5em $\underset{j} {P_V} \gets$ Protected Value
\vspace{-0.1em}
\STATE ~\hskip 1.5em $A_E$ = $|\underset{j} {P_V} - M_R|$ $\gets$ For Positive Value
\vspace{-0.7em}
\STATE \textbf{End Function}

\end{algorithmic}
\end{algorithm}


\subsubsection{Algorithmic Foundation}

Traditional smart meters directly transmit their real-time energy values to smart grid utility without using any type of privacy protection mechanism. However, our proposed model uses the advantages of dynamic differential privacy protection to protect this sensitive data. In order to demonstrate the complete technical functioning of our model we develop a detailed algorithm containing conditions for all differential privacy variants. The detailed pseudo-code is given in Algorithm~\ref{algoDP}. Similarly, the algorithm for selection of specific noise addition mechanism is given in Algorithm~\ref{algo1}.\\
The algorithm is divided into five major parts, in which the first four parts contains addition of nose via any variant of differential privacy, while the fifth part comprises of computation of absolute error value. Firstly, the input values such as instantaneous meter reading ($M_R$), number of smart meters for that specific slot ($N$), privacy budget ($\varepsilon$), sensitivity ($\delta$), database sensitivity ($\Delta$), and noise function ($F$) is fed to the algorithm to initiate the process. Afterwards, the condition for specific variant of differential privacy is checked. In case if the user calls for Laplace noise addition mechanism, then the first part ($1$) Laplace privacy mechanism is called for execution. In this part, first of all the Laplace privacy budget ($\varepsilon_L$) is taken from user. After that, database sensitivity value is computed using ~\ref{seneqn}. After computation and collection of these values, the noise is generated using the Laplace noise probability distribution function given in Eq.~\ref{lapeqn}. The values of $\varepsilon_L$ and $M_R$ are fed to the function to determine the noise scale. After successful computation of noise, this value is added to the meter reading value via $ {P_V} = {MR} + {noise}$. Next, the absolute error function is called to get the value of error, which is then stored to calculate the mean absolute error. After this process, the protected value $P_V$ is transmitted to grid utility to carry out statistical processes.\\

Similarly, if function of Gaussian noise addition is selected then the second part ($2$) of the algorithm is used to compute Gaussian noise via normal distribution. In this part, the value of Gaussian privacy budget $\varepsilon_G$ is used to compute noise in a way that it protected the final reading. The noise in this mechanism is computed using the probability density function in Eq.~\ref{gaus}. Moving further to the third part, the functionality of uniform noise addition mechanism is different from Gaussian or Laplace mechanism, because in this mechanism, the privacy budget ($\varepsilon_U$ is set to be zero, and it does not play any critical role in computation of noisy value. Contrary to this, Uniform sensitivity value of meter reading ($\delta_U$) is used to generate random noise value via uniform distribution whose equation has been provided in Eq.~\ref{uniform}. The remainder of the steps for Uniform noise addition mechanism are same as that of above two parts.\\
Similarly, if the user calls for addition of noise via Geometric differential privacy mechanism, then the fourth ($4$) part of the algorithm executes. In this part, probability of success is computed for Geometric mechanism by taking into account the value of $varepsilon_{geo}$ and $\Delta$. After successful computation of success probability ($P_R$), this value is fed to Geometric noise additions function, which then computes geometric noise according to the provided density function. Finally, the fifth part of algorithm calculates absolute error value by subtracting meter reading from the protected noise value. This function of error calculation is called for every individual perturbed value in order to keep a record error in the transmitted readings. Afterwards, these values are accumulated to calculate mean absolute error, which is demonstrated in performance evaluation section.

\section{Performance Evaluation}\label{performace}

In our blockchain based smart metering scenario, protected real-time data is transmitted from smart meters instead of original real-time data. In order to protect privacy leakage from original data, we use four different variants of differential privacy discussed in previous section. The noise produced by each variant have different effect on privacy level depending upon the privacy budget ($\varepsilon$) and reading sensitivity ($\delta$). In this section, we first discuss simulation environment and afterward provide a detailed performance evaluation on the basis of real-time reporting and mean absolute error.
\subsection{Simulation Parameters}
In order to implement differential privacy variants in our blockchain based smart metering scenario, we first of all used grid energy data from~\cite{data01} and modified it accordingly to carry out experiment for 24 hour usage. We carried out performance evaluation of differential privacy variants on above mentioned real-time smart metering data in which readings are being sent to grid utility after every 10 minutes. So, for daily usage profile, we carried out evaluation on 144 data samples collectively for each smart home. To carry out experimental evaluation, we use NumPy v1.14  and pandas v1.0.3 libraries in Python 3.0. In order to implement differential privacy variants, we use respective distribution of NumPy v1.1.4 library and modify the input parameters according to the requirement of each variant given in Section~\ref{DPVar}. \\
Furthermore, we generate the graphs at different $varepsilon$ values for three mechanism and for Uniform differential privacy, we carry out evaluation at different $\delta$ values. In this way, six graphs showing the original and protected readings for each noise adding variant are generated. Similarly, to take an account of noise and to compare efficiency of each noise addition variant, we pick the best performing $varepsilon$ value for each variant and compare their mean absolute error (MAE) values. 

\begin{figure*}[]        
\includegraphics[scale = 0.9]{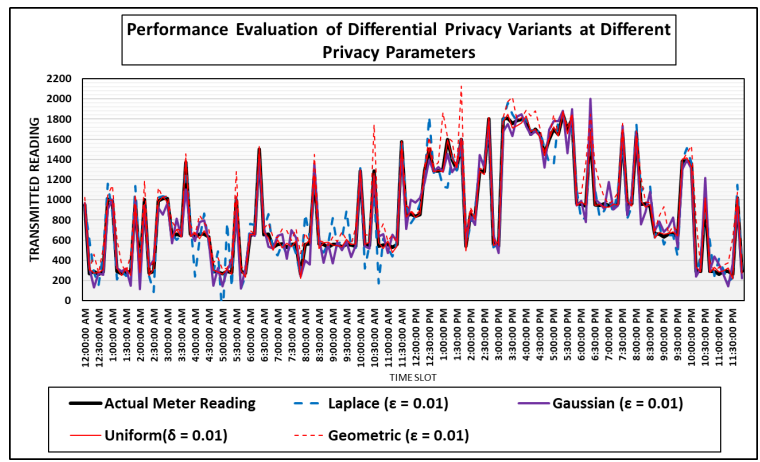}
\centering
  \caption{\small{Performance Evaluation of Differential Privacy Mechanisms at Various Privacy Parameters [Laplace $\varepsilon$ = 0.01, Gaussian $\varepsilon$ = 0.01, Uniform $\delta$ = 0.01, Geometric $\varepsilon$ = 0.01]. }}     
  \label{graph:0_01}   
\end{figure*}

\begin{figure*}[]        
\includegraphics[scale = 0.9]{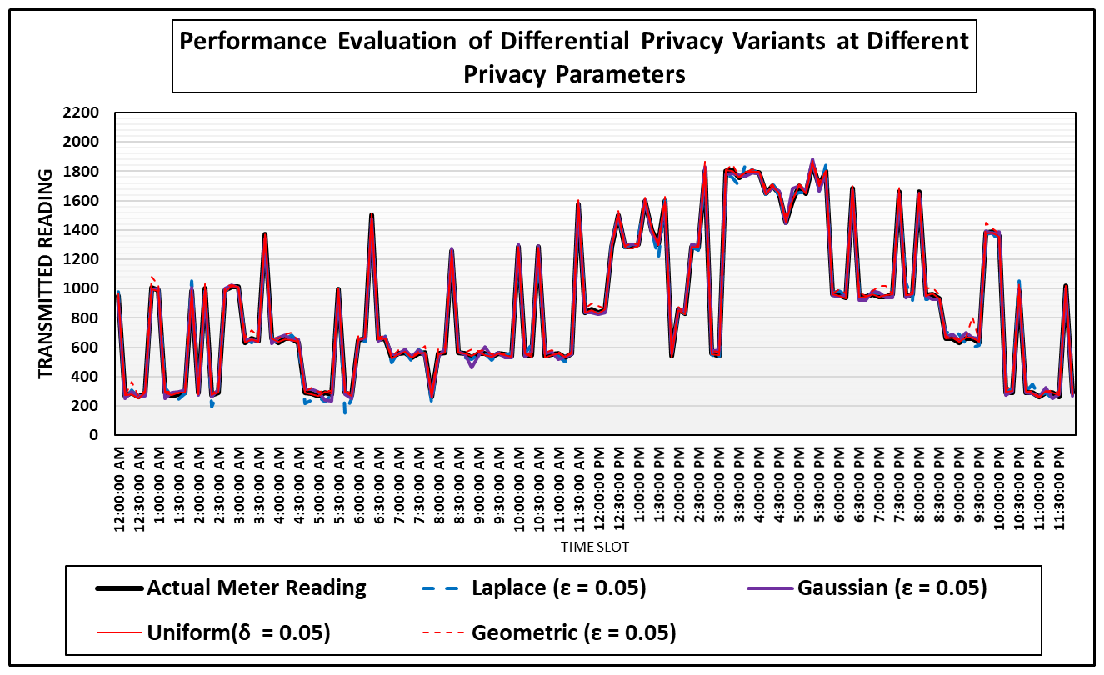}
\centering
  \caption{\small{Performance Evaluation of Differential Privacy Mechanisms at Various Privacy Parameters [Laplace $\varepsilon$ = 0.05, Gaussian $\varepsilon$ = 0.05, Uniform $\delta$ = 0.05, Geometric $\varepsilon$ = 0.05]. }}     
  \label{graph:0_05}   
\end{figure*}

\begin{figure*}[]        
\includegraphics[scale = 0.9]{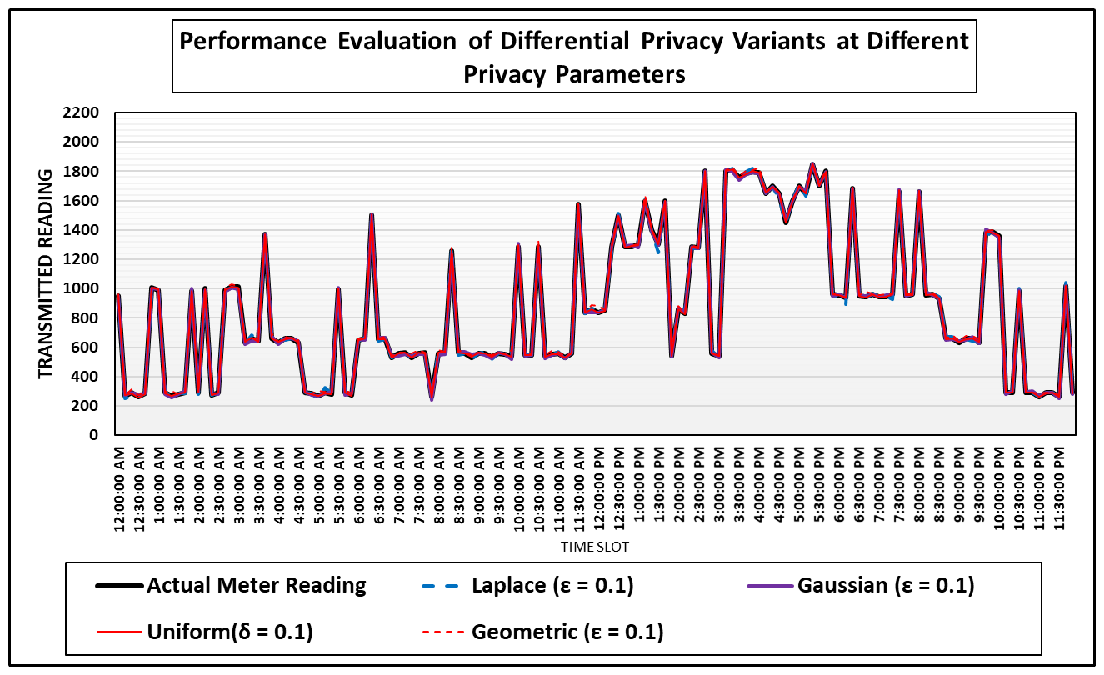}
\centering
  \caption{\small{Performance Evaluation of Differential Privacy Mechanisms at Various Privacy Parameters [Laplace $\varepsilon$ = 0.1, Gaussian $\varepsilon$ = 0.1, Uniform $\delta$ = 0.1, Geometric $\varepsilon$ = 0.1]. }}     
  \label{graph:0_1}   
\end{figure*}

\begin{figure*}[]        
\includegraphics[scale = 0.9]{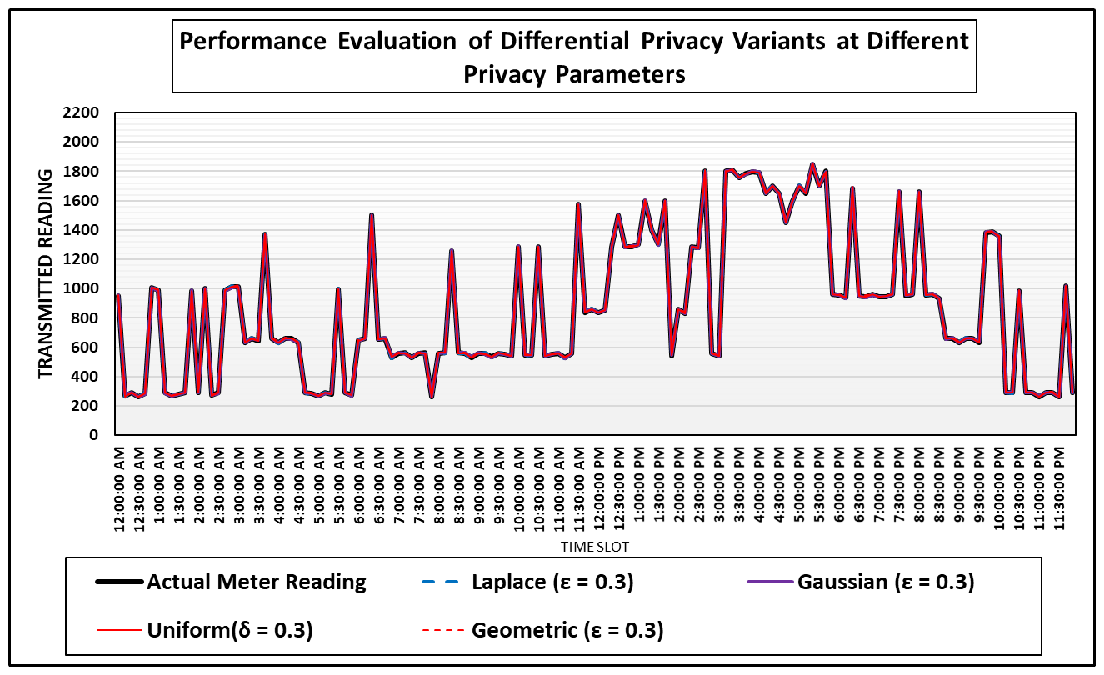}
\centering
  \caption{\small{Performance Evaluation of Differential Privacy Mechanisms at Various Privacy Parameters [Laplace $\varepsilon$ = 0.3, Gaussian $\varepsilon$ = 0.3, Uniform $\delta$ = 0.3, Geometric $\varepsilon$ = 0.3]. }}     
  \label{graph:0_3}   
\end{figure*}

\begin{figure*}[]        
\includegraphics[scale = 0.9]{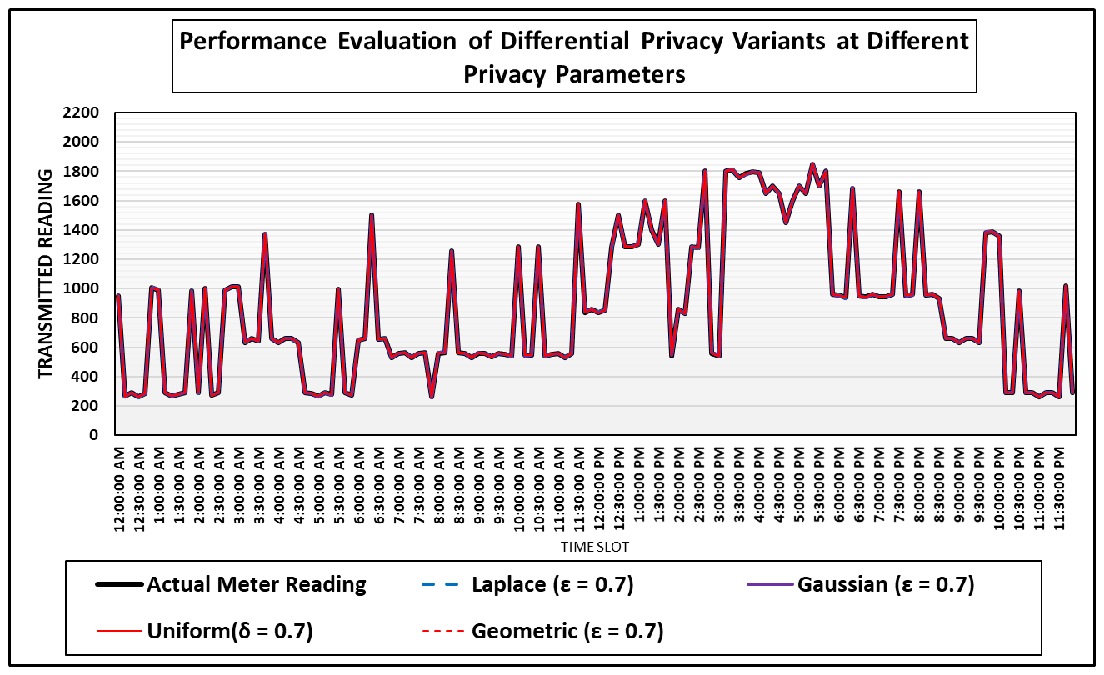}
\centering
  \caption{\small{Performance Evaluation of Differential Privacy Mechanisms at Various Privacy Parameters [Laplace $\varepsilon$ = 0.7, Gaussian $\varepsilon$ = 0.7, Uniform $\delta$ = 0.7, Geometric $\varepsilon$ = 0.7]. }}     
  \label{graph:0_7}   
\end{figure*}

\begin{figure*}[]        
\includegraphics[scale = 0.9]{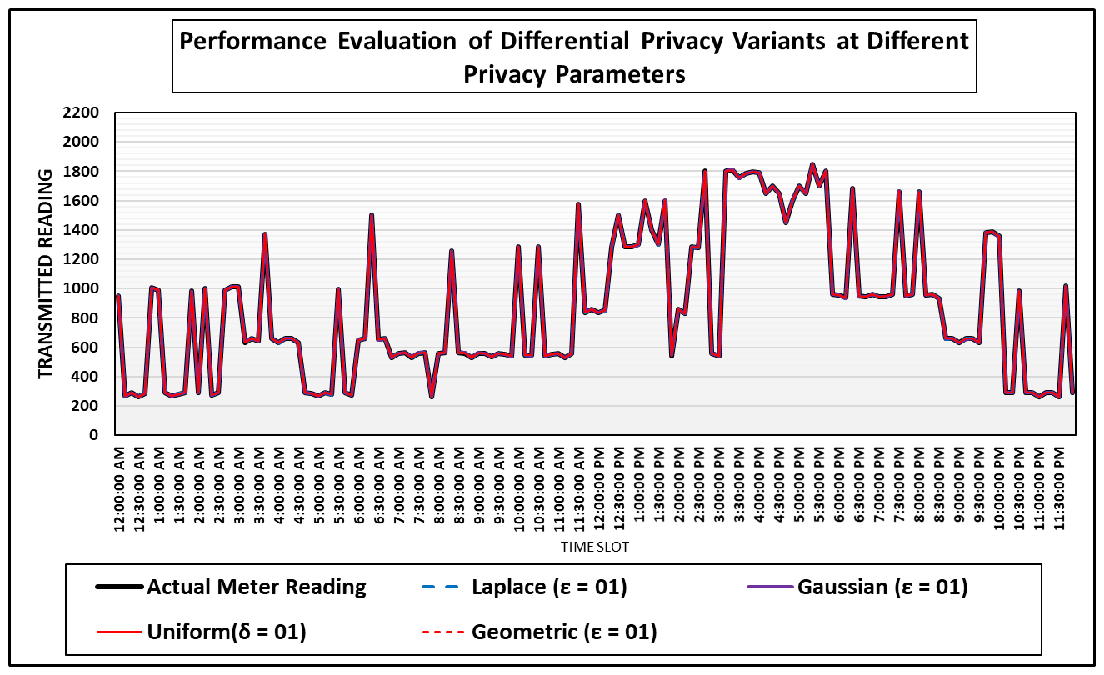}
\centering
  \caption{\small{Performance Evaluation of Differential Privacy Mechanisms at Various Privacy Parameters [Laplace $\varepsilon$ = 1, Gaussian $\varepsilon$ = 1, Uniform $\delta$ = 1, Geometric $\varepsilon$ = 1]. }}     
  \label{graph:1}   
\end{figure*}


\subsection{Private Real-Time Data Reporting}
Reporting real-time smart meter data can leak users’ privacy and can cause serious threats to occupants of smart homes, therefore, we use differential privacy to protect this information. In order to do so, we implement four variants and analysed their output in the form of given graphs. In Fig.~\ref{graph:0_01}, the graph for $\varepsilon$ = 0.01 can be seen for Laplace, Gaussian, and Geometric noise addition mechanism. Similarly, the graph for $\delta$ = 0.01 for Uniform noise addition mechanism is also plotted in the same figure for comparison purpose. The graph depicts values transmitted in 24 hours by using different variants as compared to the actual meter reading. Starting from first quarter of the graph, ranging from 12:00AM to 06:00AM,  the usage is pretty low because of night, and only few peaks can be seen. However, the behaviour of all noise addition mechanism can be observed clearly. If one closely analyses the graph, it can easily be seen that Laplace mechanism provided more diversified amount of noise as compared to any other noise addition mechanism in this quarter. After Laplace, some variation for Gaussian mechanism can also been for some low peaks such as 04:30AM, and after that some at some high peak values Geometric mechanism shows variance as compared to other three. \\
Moving further to second quarter of the graph from 06:00AM to 12:00PM, similar trend can be observed, that whenever there is low value of original meter reading, then Laplace noise addition mechanism provides maximum variance followed by Gaussian mechanism, which usually provides negative variance as compared to the trend. And whenever there is high peak value, such as 10:30AM when the original usage is around 1200Wh, then the Geometric mechanism add a considerable amount of noise to protect privacy of real-time smart metering data. After that, in the third quarter ( which is the most highly utilised quarter), Geometric noise addition mechanism outperforms other variants, because of its usual reputation of protecting reading at higher peaks. At it can be seen that around 01:00PM, the value of load usage is pretty high around 1600Wh, and at this stage Geometric mechanism provides highest peak value by providing maximum fluctuation. This trend continues and Geometric mechanism keeps on dominating other mechanism because the usage values are pretty high. After Geometric, the second most dominant noise addition mechanism here is Laplace mechanism, which protect the peak values by either adding negative noise or either by adding minor positive noise depending upon the noise calculation. However, the result of remaining two mechanisms can partially be seen at this stage.\\
In the fourth quarter of the graph from 06:00PM to 12:00AM, the load utilization again reduces, and in this quarter, the trend similar to quarter number 2 can be seen. Where Laplace mechanism dominates because of low peak values and after that Gaussian mechanism came into light because of the negative random noise it is adding in those low peak values. The variation because of Geometric mechanism can also be seen at some places, although, it is not as dominant as that of Laplace and Gaussian mechanism. Throughout the graph, Uniform noise addition mechanism does not came into lime-light which is because of the reason that its noise span is pretty low as compared to other three mechanisms. If one zooms the graph, it surely can see the variations because of Uniform mechanism, but due to high variations by other three variants, it gets supressed.\\
A similar trend can be seen for remaining output graphs as well. For example, in Fig.~\ref{graph:0_05}, the graph for $\varepsilon$ = 0.01 for Laplace, Gaussian, and Geometric mechanism. And $\delta$ = 0.01 for Uniform noise addition mechanism can be seen. As the value of privacy budget is increased, which means less privacy protection and more utility, so it can be observed that at majority of places, very minimal noise has been added to protect users’ privacy. Some low peaks can be seen for Laplace mechanism in the graph at the time of less usage, however, majority of added noise values remain unnoticed because of not producing much variation. Moving further to other graphs presented in Fig. ~\ref{graph:0_1}, ~\ref{graph:0_3}, and ~\ref{graph:0_7}, the variation reduces, and very minimal noise addition can be observed. Finally, in Fig.~\ref{graph:1}, when the value of $varepsilon$ and $\delta$ is increased to ‘1’, the added noise almost reaches zero, which means that minimum privacy and maximum utility. At this privacy budget, mostly the actual meter is reported as it is to the grid utility with very minimal and negligible variation because of noise. Therefore, it can be concluded that lower values of privacy parameters provide high privacy protection, and increasing these values result in gradual decrease in privacy protection to a limit that utility becomes maximum with very minimal privacy protection as demonstrate in the graphical figures. \\
Keeping in view all the graphs and discussion, it can be said that in high peak values, Geometric mechanism provides healthy variation to protect privacy, and for low peak values, Laplace mechanism followed by Gaussian mechanism provides considerable variation to protect users’ privacy.

\begin{figure*}[]        
\includegraphics[scale = 1.5]{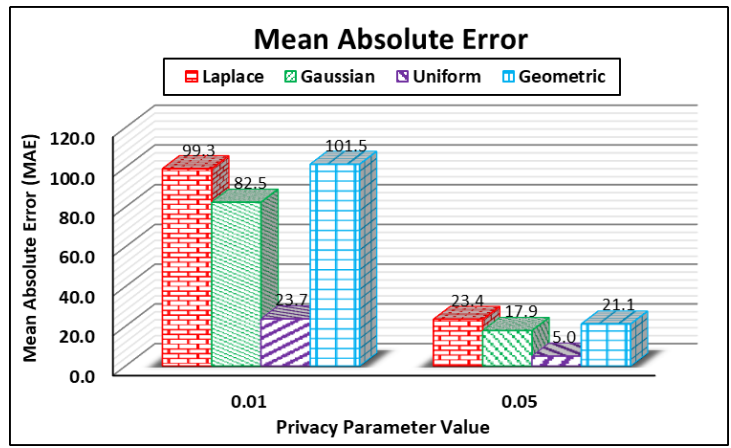}
\centering
  \caption{\small{Mean Absolute Error of Differential Privacy Variants at Different Privacy Parameters }}     
  \label{graph:mae}   
\end{figure*}

\subsection{Mean Absolute Error}

In order to provide statistical analysis regarding differential privacy noise addition variants, we use the parameter of MAE. Formally, MAE can be termed as the difference between the noisy value and the original reading. The equation for MAE is given as follows:
\begin{equation}
\centering
MAE = \sum_{n=1}^{N_R} |P_v – M_R|
\end{equation}

In the above equation, $N_R$ is the total number of noisy readings sent in the day. In our mechanism, we accumulate energy usage of every 10 minutes and then transmit the values to grid utility after addition of noise. Therefore, the value of $N_R$ = 144, although, this value can vary according to the dataset. Similarly, $P_v$ is the protected noisy value, and $M_R$ is the actual meter reading which demonstrate the actual usage within 10 minutes by smart meter user. A graphical illustration of MAE values is given in Fig.~\ref{graph:mae}.\\
In Fig.~\ref{graph:mae}, we show graphical illustration at $\varepsilon, delta$ = 0.01 and 0.05, because these two privacy parameter values add an adequate amount of noise to protect privacy. After further increment of $\varepsilon$ or $\delta$, this value of MAE decreases and eventually reduces to approximately zero at $\varepsilon, \delta$ = 1. Furthermore, the average meter reading value for 144 readings in daily load is 872Wh per reading, which means that approximately 872Wh is being used every 10 minutes. \\
It can be seen from the graph that Geometric noise addition mechanism provides MAE value of 101.5 at $\varepsilon$ = 0.01, followed by Laplace and Gaussian mechanism who provides MAE of 99.3 and 82.5, respectively. Similarly, at $\varepsilon$ = 0.05, the MAE value 23.4 provided by Laplace is maximum, which is followed by Geometric and Gaussian noise addition mechanism as 21.1 and 17.9, respectively. From Fig.~\ref{graph:mae}, it can be concluded that both Geometric and Laplace mechanisms provides similar MAE, however, from the above section, it can be seen that Laplace provides efficient variation in case of small meter reading and Geometric provides variation in case of large used values.
\subsection{Summary \& Lessons Learnt}
After careful visualization of real-time smart metering graphs and MAE graph, it can be concluded that both Laplace and Geometric mechanisms outperform other noise addition variants by providing more variation as compared to other two. However, if the data usage values are large, then geometric noise addition is more suitable because it provides more variation in case of large values by showing high positive peaks. Contrary to this, Laplace mechanism usually provides negative peaks in case of low values to protect user privacy. Furthermore, from perspective of $\varepsilon \& \delta$, the lower these values are, the higher the privacy is, therefore, in order to protect at least an adequate amount of privacy, $\varepsilon, \delta$ = 0.01 are the most suitable privacy parameters. Furthermore, MAE evaluation value of Geometric and Laplace mechanism are quite similar, which also shows that both of the mechanism are efficient to protect smart metering privacy. Though, if there is a smart home in which on average less energy is used, then Laplace mechanism (at $\varepsilon$ = 0.01) outperforms other privacy preserving mechanisms, and in case if there is a smart home which regularly uses high amount of energy throughout the day, then Geometric noise addition mechanism (with $\varepsilon$ = 0.01) is the most feasible one to be used.

\section{Conclusion and Future Directions}

Differential privacy appeared as a strong privacy preserving notion after its invent by C. Dwork in 2006. Since then, plenty of variants of differential privacy have been proposed by researchers that have been applied over certain real-time application, and real-time smart metering is one of them.  Similarly, the security and transparency of smart metering has also been enhanced by usage of decentralized blockchain technology. In this chapter, we first work over integration of differential privacy and blockchain in real-time smart metering scenario. Afterwards, we carried out performance evaluation of four variants of differential privacy in the proposed blockchain based smart metering scenario. Performance evaluation section of the chapter demonstrate that each privacy preserving mechanism adds performs differently depending upon the selected privacy parameters and the input data. However, in case of high peak values Geometric mechanism surpasses other noise addition variant, although, in case of low peak values in meter reading, Laplace mechanism performs better at $\varepsilon$ = 0.01. As a part of future work, we are working over integration of differential privacy  and blockchain in other cyber physical systems scenarios.

\bibliographystyle{IEEEtran}

\

\end{document}